\documentclass{article}
\usepackage{spconf,amsmath,graphicx,color}
\usepackage{multirow}

\usepackage{lineno}

\title{SingNet: A Real-time Singing Voice Beat and Downbeat Tracking System}
\name{\hspace{-0.5in}Mojtaba Heydari$^{\ddagger \dagger}$ \qquad Ju-Chiang Wang $^{\star}$ \qquad Zhiyao Duan$^{\dagger}$ \thanks{$\ddagger$ Main work completed as a research intern at TikTok
Inc. This work is partially funded by National Science Foundation grant
No. 1846184. }}

\address{$\hspace{-0.3in}^{\dagger}$ University of Rochester, Rochester, NY, USA \hspace{0.5in}
      $^{\star}$ TikTok, Mountain View, CA, USA\\ \tt \normalsize \hspace{-0.18in} mheydari@ur.rochester.edu  \hspace{0.1in}  ju-chiang.wang@tiktok.com \hspace{0.1in} zhiyao.duan@rochester.edu} 

\begin{document}
\ninept

\maketitle

\begin{abstract}
Singing voice beat and downbeat tracking posses several applications in automatic music production, analysis and manipulation. Among them, some require real-time processing, such as live performance processing and auto-accompaniment for singing inputs. This task is challenging owing to the non-trivial rhythmic and harmonic patterns in singing signals. For real-time processing, it introduces further constraints such as inaccessibility to future data and the impossibility to correct the previous results that are inconsistent with the latter ones. In this paper, we introduce the first system that tracks the beats and downbeats of singing voices in real-time. Specifically, we propose a novel dynamic particle filtering approach that incorporates offline historical data to correct the online inference by using a variable number of particles. We evaluate the performance on two datasets: GTZAN with the separated vocal tracks, and an in-house dataset with the original vocal stems. Experimental result demonstrates that our proposed approach outperforms the baseline by 3--5\%.

\end{abstract}
\section{Introduction}
\label{sec:intro}

Music rhythmic analysis is an essential Music Information Retrieval (MIR) task with many applications. Recent years have seen several deep learning-based models proposed for music beat, downbeat, tempo, and meter tracking. For instance, some employ Recurrent Neural Networks (RNN) to model beats and downbeats in music audio \cite{bock2016madmom, krebs2016downbeat,fiocchi2018beat,chiu2021drum}. Another set of works are based on Convolutional Neural Networks (CNN)~\cite{di2021downbeat,schreiber2018single}, Convolutional Recurrent Neural Networks (CRNN)~\cite{fuentes2019music} and Temporal Convolutional Networks (TCN)~\cite{matthewdavies2019temporal}. More recently, some works take advantage of Transformers ~\cite{hung2022modeling,zhao2022beat} and self-supervised learning~\cite{desblancs2022self} to improve the performance. All of the mentioned methods are offline and mostly utilize a non-causal Dynamic Bayesian Network (DBN) decoder to infer music beats and downbeats. On the other hand, a group of works are capable of performing causally for real-time applications. Examples such as~\cite{gkiokas2017convolutional,bock2011enhanced} applied a sliding window technique on an offline model to predict the upcoming beats. Main disadvantages of this technique include the potential for computational overload and the lack of continuity between windows~\cite{oliveira2010ibt}. Lately, particle filtering mechanism in conjunction with the RNN and CRNN structures is proposed to estimate the rhythmic parameters in an online fashion~\cite{heydari2021don,heydari2021beatnet}. This combination makes the inference inherently causal.

Recently, beat and downbeat tracking for isolated singing voices was proposed \cite{heydari2022singing} to fulfill the requirements for applications such as automatic music arrangement, remix, and sing-to-song, where a user wants to create a new song starting with an arbitrary acapella singing melody.
However, isolated singing voices are considered to be more challenging than full music signals (i.e., the mixture of vocal and accompaniment) for rhythmic analysis, because they typically lack rich percussive and harmonic profiles.
As demonstrated in \cite{heydari2022singing}, 
applying general music rhythmic analysis approaches to singing voice beat and downbeat tracking is less effective. In the same work, 
two models are proposed to tackle offline singing beat tracking. The proposed models employ pre-trained WavLM~\cite{chen2022wavlm} and DistilHuBERT~\cite{chang2022distilhubert} speech representations respectively as their front-end to leverage semantic information of singing voices. In addition, they use multi-head linear self-attention encoder layers ~\cite{katharopoulos_et_al_2020}, followed by an offline DBN decoder, to 
infer the singing beats.

For many use cases such as interactive social media gadget design, singing voice auto-accompaniment, and live performance processing, real-time singing voice beat and downbeat tracking is essential. Real-time processing imposes causality constraints to the system, including partial data accessibility and having no second chance to correct the previously inferred results according to the new data. Also, the neural network structures need to be causal and computationally efficient. Therefore, it limits the adequacy of bulky Transformers and massive pre-trained speech models which are previously demonstrated to be beneficial for offline singing voice beat tracking.   

\begin{figure*}[ht]
 \centerline{
 \includegraphics[width=2\columnwidth]{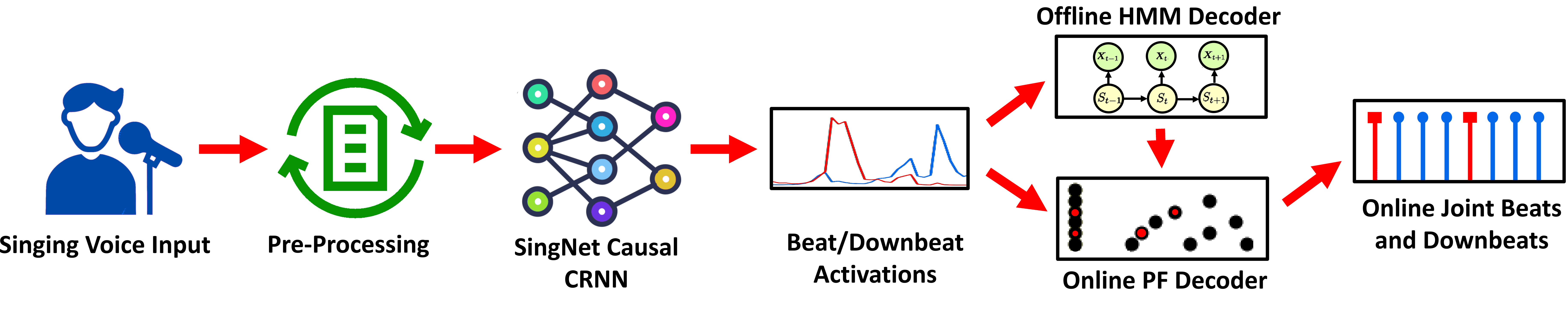}}
 \caption{Overall pipeline of the SingNet system.}
 \label{overview}
\end{figure*}

In this paper, we introduce SingNet, the first online joint beat and downbeat tracking system for singing voices. It is noted that the term ``singing voices'' specifically refer to the isolated singing voices without instrumental accompaniments. SingNet adopts a CRNN to model the beat and downbeat activations. On top of that, we propose a novel dynamic particle filtering model that leverages a variable number of particles for the inference instead of the fixed number of particles used in traditional particle filtering approaches. The proposed dynamic particle filtering incorporates the offline inference results on all historical data into the ongoing online inference process by manipulating the number and positions of the particles. Because rhythm analysis of singing voices is more robust when taking into account the past signal, such analysis results can be a informative prior to improve the online inference. Furthermore, to take all substantial activations into account, it adds extra particles when there is a considerable salience happening. 

The rest of the paper is organized as follows. Section 2 elaborates the proposed
approach. In Section 3, we present the experimental results with comparisons to those of relevant methods. Finally, section 4 concludes the paper.

\section{Proposed Approach}
\label{sec:approach}

In this section, 
we first describe the neural network structure, and then elaborate the proposed dynamic particle filtering inference approach.
Figure \ref{overview} depicts the general diagram of the SingNet system. 

\subsection{Model}

Music rhythmic analysis approaches usually use acoustic features such as magnitude spectrogram as their input, while isolated vocal tracks are more similar to speech signals where the para-linguistics, semantic, and phonemic level features are important as well~\cite{heydari2022singing}. Using the features generated from pre-trained speech models together with a multi-head linear Transformer is proposed to tackle the offline singing voice beat tracking~\cite{heydari2022singing}.
However, as the mentioned models are non-causal and computationally expensive, they cannot be used for the current task.
Therefore, in this work, we focus on the conventional spectral features with a regular causal model similar to BeatNet~\cite{heydari2021beatnet}. 

To define SingNet's neural structure, we start off with the causal CRNN as the backbone. Based on our pilot experiments with various modifications, we expand the model used in BeatNet in both depth and width to have 3 unidirectional LSTM layers, and each layer has 350 hidden cells. The resulting improvement 
is in line with our hypothesis: the rhythmic analysis of isolated singing voices needs to deal with more non-trivial components as compared to the music pieces of the full mixture, where the harmonic and percussive patterns are clearer. Therefore, using a larger model with higher complexity is beneficial. Finally, the model outputs the beat, downbeat, and non-beat activations for each time frame, which will be fed to the inference model later. 

\begin{figure}[!b]
 \centerline{
 \includegraphics[width=1\columnwidth]{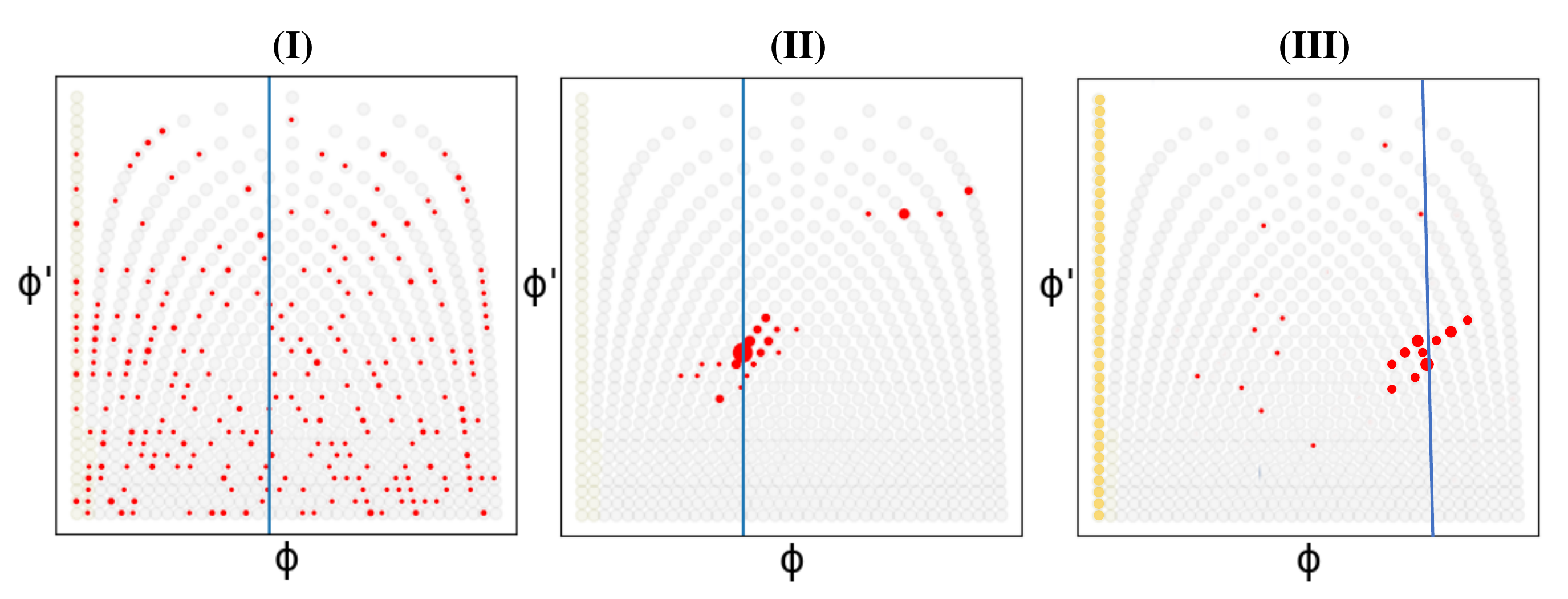}}
 \caption{The particle filtering process of the beat/tempo tracking. The gray dots are all possible states, $\phi$ and $\phi^\prime$ represent the phase and tempo axes of the state, red dots are the particles, and the vertical blue line is their median along phase. (I) corresponds to the beginning of the process where the particles are distributed randomly. (II) displays one of the later time steps when the particles are converged to a swarm which is moving forward according to the transition model. (III) presents a time step when a salience activation (in yellow) occurs in the beat states (i.e., the first column in the state space), but the particle swarm is in a distant position to the beat states.
 }
 \label{inference}
\end{figure}

\subsection{Inference}

The inference model consists of a cascade of two-stage Monte Carlo particle filtering,
one for the beat and the other for downbeat tracking.
The inference state space, observation and transition models are the same as those of BeatNet.
However, instead of using fixed number of particles in conventional models~\cite{heydari2021don,heydari2021beatnet,hainsworth2004particle}, we introduce a dynamic particle filtering technique, which leverages a variable number of particles depending on the circumstance. Figure \ref{inference} is an example of a 2D beat pointer state space~\cite{krebs2015efficient} and the particle filtering inference on it.

We propose three schemes: \emph{salience-informed}, \emph{past-informed}, and \emph{combined} methods, to incorporate the historical data derived from an offline inference into the online decoding as well as to prevent the model from omitting predominant activations during the inference process.

The \emph{salience-informed} method takes into account the instant saliences to gain robustness to local rhythmic fluctuation/change. The saliences are defined by thresholding the activations with 0.4 in our implementation. In regular cases, the particle swarm returns to the beat/downbeat position at the same time when the next predominant activation is happening. However, as shown in Figure~\ref{inference}-(III), there might be some inconsistency that the time steps of particles are far from the beat/downbeat states when a strong beat/downbeat activation is occurring. This can be due to either a tempo change or an inference mistake. In other words, the inference is not responsive to such salient activations.
To prevent that, the method adds extra particles at beat/downbeat states when a strong activation is detected.

\begin{figure}[t]
 \centerline{
 \includegraphics[width=1\columnwidth]{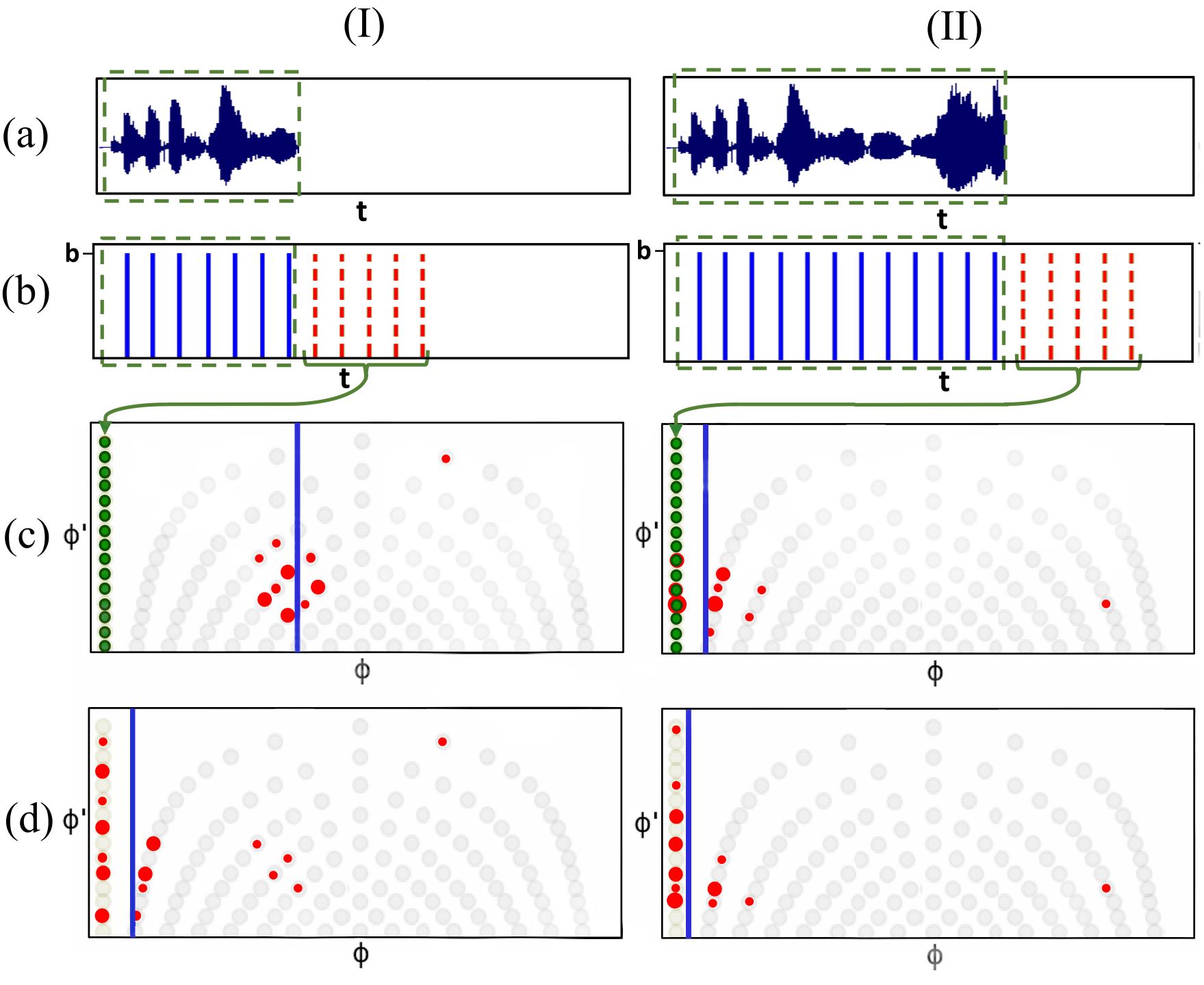}}
 \caption{An example of \emph{past-informed} process for two time steps (I) and (II). (a) the arriving streaming audio; (b) solid blue lines are historical beats inferred by the offline DBN on past activations, and dotted red lines are the extrapolated results from blue lines. (c) injecting new particles (in green) into the beat/tempo state space before resampling, based on the extrapolations. (d) the phase correction after resampling.}
 \label{past-informed}
\end{figure}

The \emph{past-informed} method comprises a fusion of simultaneous online and offline inferences. Given a period parameter $T$, the DBN decoder is applied every $T$ seconds to the past data, i.e., entire activations from the beginning to the present time step. The historical beat/downbeat timestamps are used to extrapolate the next upcoming beats/downbeats. When the time steps of the upcoming beats/downbeats arrive, the method injects some particles at the beat/downbeat states to `inform' the online inference about the offline extrapolations. 
If the particle swarm locates in some offbeat states but the offline inference suggests a beat/downbeat onset, adding the extra particles can help correct the online inference. Figure \ref{past-informed} demonstrates the process. To save space, we only show the beat/tempo process, and the downbeat/meter process is the same. At time step (I), since there is a big discrepancy between the offline extrapolations and the online particle swarm (i.e., the median blue line in (c) is far from the beat state), adding particles shifts the beat phase significantly. Rather, at time step (II), as the online beat status is already corrected, adding particles does not change the beat phase much.

The \emph{combined} method combines \emph{salience-informed} and \emph{past-informed}  methods by intersecting the particle injections of the above two methods.
For all of the mentioned methods, the same number of particles to the injected ones are randomly selected and removed from the original particle pool after the resampling step to ensure the total number of particles is the same for every iteration throughout the whole inference process.








\section{Experiments}
\label{Experiments}

\subsection{Datasets and Evaluation Metrics}

Evaluating the beat and downbeat tracking for singing voices is challenging. To our knowledge, there is no existing public datasets containing clean vocal audio with beat/downbeat annotations. Previous work has found that annotating beats and downbeats on sole vocal signals even by human can be difficult and subjective \cite{heydari2022singing}, since no explicit rhythmic clues such as percussive instruments can be referred to truly understand the singer's rhythmic intentions. Therefore, music source separation (MSS) techniques \cite{rafii2018overview} were used to separate the vocal stems from the original mixture audio, and the beat/downbeat annotations applied on the mixture audio are adopted to describe the true rhythmic intentions.

To evaluate our system, we used \emph{GTZAN}~\cite{marchand2015swing} which is a publicly available dataset and was not used during training, as described in~\cite{heydari2022singing}.
In \emph{GTZAN}, 741 clips contain vocals, so we utilized Demucs Hybrid~\cite{defossez2021hybrid} to extract the vocal stems for evaluation. However, MSS can result in leaks of instrument sounds, which would provide clues for the model to exploit. To this end, we also used an \emph{In-House} dataset, which contains about 45,000 music clips of clean vocal stems as well as accompaniments with beat and downbeat annotations. Each clip is excerpted from an individual cover song. Song examples of the \emph{In-House} dataset may be found online.\footnote{https://www.karaoke-version.com/} We split the dataset into 43,000 clips for training, 1000 for validation, and 1000 for testing respectively. The testing set is fully unseen during the training. The trained model on \emph{In-House} is evaluated on \emph{GTZAN} in a cross-dataset evaluation setting as well. 


Following the typical evaluation settings \cite{heydari2021don}, we report the F1 scores with a tolerance of $\pm70$ ms with and without 5-sec skip from the beginning of a test clip. 
As mentioned above, annotating beats and downbeats for pure singing voices can be subjective. Based on our pilot subjective study, we observed that human tolerance to beat and downbeat errors can be larger compared to regular music signals, and the adequate time range is about $\pm200$ ms. As a result, we also report the F1 scores of $\pm200$ ms tolerance.


\subsection{Implementation Details}

To train the SingNet neural network, we initialized all weights and biases
randomly and trained the model with Adam optimizer, weighted cross entropy loss, learning rate of $5 \times 10^{-5}$ and a batch size of 50. Every Batch comprises 8-second long excerpts randomly sampled from each track. To have a more robust model, we augmented the training samples ~\cite{spijkervet_torchaudio_augmentations}. 

We set $T = 6$ seconds for the aforementioned period parameter. The choice of applying the DBN every 6 seconds is motivated by the prior works of conventional beat/downbeat tracking (e.g., \cite{bock2016madmom}), where they typically adopted 6 second for the input window length. In our pilot study, we did not find substantial differences when significantly increasing or decreasing the 6 seconds period parameter.


\renewcommand{\arraystretch}{1.2}
\begin{table*}[t]
  \begin{center}
  \resizebox{500pt}{!}{
\begin{tabular}{l|cccc|cccc}
 & \multicolumn{4}{c|}{No Skip} & \multicolumn{4}{c}{Skip First 5 Seconds} \\[3pt]
Metric  & 
\renewcommand{\arraystretch}{0.8}\begin{tabular}[c]{@{}c@{}}Beat \\ (70ms)\end{tabular} & 
\renewcommand{\arraystretch}{0.8}\begin{tabular}[c]{@{}c@{}}Downbeat \\  (70ms)\end{tabular} & 
\renewcommand{\arraystretch}{0.8}\begin{tabular}[c]{@{}c@{}}Beat \\ (200ms)\end{tabular} & 
\renewcommand{\arraystretch}{0.8}\begin{tabular}[c]{@{}c@{}}Downbeat \\  (200ms)\end{tabular} & 
\renewcommand{\arraystretch}{0.8}\begin{tabular}[c]{@{}c@{}}Beat \\ (70ms) \end{tabular} & 
\renewcommand{\arraystretch}{0.8}\begin{tabular}[c]{@{}c@{}}Downbeat \\  (70ms) \end{tabular} & 
\renewcommand{\arraystretch}{0.8}\begin{tabular}[c]{@{}c@{}}Beat \\ (200ms) \end{tabular} & 
\renewcommand{\arraystretch}{0.8}\begin{tabular}[c]{@{}c@{}}Downbeat \\ (200ms) \end{tabular} 
\\ 
\hline
\multicolumn{9}{c}{\normalsize\emph{In-House Dataset}} \\
\hline
BeatNet (baseline)                   & 53.90                                                         & 30.06                                                             & 71.40                                                         & 42.33                                                              & 54.90                                                                & 31.58                                                                     & 71.97                                                                 & 43.90                                                                     \\ \hline
SingNet (default inference)       & 56.48                                                     & 34.36                                                         & 72.84                                                     & 46.25                                                          & 57.57                                                                & 36.55                                                                     & 73.64                                                                 & 48.43                                                                     \\ \hline
SingNet (salience-informed) & 58.09                                                     & 35.74                                                         & 74.54                                                     & 47.23                                                          & 59.51                                                                & 37.99                                                                     & 75.53                                                                 & 49.22                                                                     \\ \hline
SingNet (past-informed) & 57.38                                                     & 35.06                                                         & 73.78                                                     & 46.80                                                          & 58.86                                                                & 37.59                                                                     & 74.74                                                                 & 48.90                                                                    \\ \hline

SingNet (combined) & \textbf{58.1}                                             & \textbf{35.95}                                                & \textbf{74.62}                                            & \textbf{47.38}                                                 & \textbf{59.53}                                                                & \textbf{38.04}                                                                     & \textbf{75.65}                                                                 & \textbf{49.24}                                                                     \\ \hline
Online DBN                 & 38.94                                                         & 20.16                                                            & 55.32                                                         & 27.14                                                              & 40.83                                                                & 20.73                                                                     & 55.20                                                                 & 27.90                                                                     \\ \hline   \hline
Offline DBN (oracle)               & 60.67                                                     & 45.17                                                         & 79.70                                                     & 55.11                                                          & 60.81                                                                & 45.22                                                                     & 79.91                                                                 & 55.20                                                                     \\ 

\hline
\multicolumn{9}{c}{} \\
\multicolumn{9}{c}{\normalsize\emph{GTZAN Dataset}} \\ 

\hline
BeatNet (baseline)                & 32.64                                                         & 14.17                                                             & 55.39                                                         & 25.45                                                              & 32.76                                                                & 14.55                                                                     & 53.96                                                                 & 25.15                                                                     \\ \hline
SingNet (default inference)       & 34.43                                                     & 16.52                                                         & 55.81                                                     & 27.97                                                          & 34.68                                                                & 17.45                                                                     & 54.52                                                                 & 27.80                                                                     \\ \hline
SingNet (salience-informed) & 34.91                                                     & 16.91                                                         & 56.39                                                     & 27.94                                                          & 35.30                                                                & 17.14                                                                     & 55.05                                                                 & 27.49                                                                     \\ \hline
SingNet (past-informed) & 34.67                                                     & 16.77                                                         & 55.97                                                     & 27.92                                                          & 35.11                                                                & 17.61                                                                     & 54.73                                                                 & 28.50                                                                     \\ \hline
SingNet (combined) & \textbf{35.14}                                            & \textbf{17.01}                                                & \textbf{56.60}                                            & \textbf{28.02}                                                 & \textbf{35.38}                                                                & \textbf{18.09}                                                                     & \textbf{55.50}                                                                 & \textbf{28.65}                                                                     \\ \hline
Online DBN                 & 22.02                                                         & 9.34                                                            & 40.70                                                         & 16.87                                                              & 24.43                                                                & 9.97                                                                     & 45.17                                                                 & 18.00                                                                     \\ \hline   \hline
Offline DBN (oracle)               & 38.29                                                     & 23.04                                                         & 64.09                                                     & 33.65                                                          & 38.55                                                                & 23.83                                                                     & 60.93                                                                 & 32.21                                                                     \\ \hline
\end{tabular}%
}
    \caption{Evaluation results (F1 scores in \%) of different methods of SingNet and comparing them to the baseline models.}
    \label{tab:results}
  \end{center}
\end{table*}

\subsection{Methods Compared}

We use BeatNet~\cite{heydari2021beatnet} as our baseline, which is trained on singing stems and uses default particle filtering. BeatNet and SingNet structures differ in the number of LSTM layers and hidden cells. BeatNet has two layers with 150 cells, while SingNet has three layers with 350 cells. We test SingNet with default and three dynamic particle filtering variants (salience-informed, past-informed, and combined).

Other than that, we also implement the ``Online DBN'' method, which uses the extrapolated results from an offline DBN applied every 6 seconds on total historical data~\cite{krebs2015efficient} for the causal uses. For all the above mentioned online methods that use a DBN inference, we apply it firstly right after 5 seconds, and every 6 seconds thereafter. 

Finally, we include the ``Offline DBN'' method, which represents an oracle system that applies the offline DBN on the entire data, assuming it can access the future signal.

\subsection{Results and Discussions}
\label{Results}
Table \ref{tab:results} presents the evaluation results of different methods. The training, validation, and test sets remain the same across all of the methods. 



Several observations are made. First, SingNet with dynamic particle filtering (i.e., salience-informed, past-informed, and combined) outperforms that using the default particle filtering, and among the variants, the \emph{combined} method performs the best in all cases. 
Although the improvement by the past-informed method is relatively smaller compared to the salience-informed method, it contributes to the combined method that leads to the best results.
Second, ``Online DBN'' results in the worst performance, since the extrapolations are not responsive to instant rhythmic fluctuations. 
Third, ``Offline DNB'' can be seen as the upper-bound performance. Causality conditions degrade all online results to be lower than those of the offline model. 
Fourth, our models perform significantly better on \emph{In-House} than on \emph{GTZAN}. Other than the data similarity between the training and testing sets, it is likely due to the artefacts and accompaniment residuals in \emph{GTZAN} caused by music source separation. Since our models are trained on clean vocal stems, it is believed that they do not take advantage of the data leakage. Consequently, the results of \emph{In-House} reflect the true performance when utilizing SingNet for user singing or humming.

However, we consider that our models still suffer from low performance of online downbeat tracking.
Nevertheless, online downbeat tracking is a challenging MIR task, with the state-of-the-art F1 score being around 50\% on the \emph{GTZAN} dataset for regular music (not singing voice only)~\cite{heydari2021beatnet}. When it comes to online downbeat tracking of isolated singing tracks, it becomes even more challenging, leading to lower results.     

In terms of overall computational complexity, the DBN used in the past-informed and combined methods does not add too much cost, as it is called rarely (once per 300 frames), and it replies on the same activations as the online inference model does. More importantly, since DBN is used to provide extrapolations for future predictions instead of immediate results, it can run in the background in parallel with other post-processing modules. Therefore, there is no perceptual delay in our implementation with $T=6$ seconds.

While we acknowledge that online methods, including the one proposed here, may not perform as well as our previously proposed offline singing voice rhythmic analysis models (reported in~\cite{heydari2022singing}), they have the advantage of being able to analyze streaming audio in real-time. The offline methods utilized large self-supervised pre-trained speech models to extract features and Transformer architectures to model the beat activation, resulting in significant improvements over the baseline methods. For instance, Table 2 of~\cite{heydari2022singing} demonstrates that replacing conventional spectral inputs with the WavLM embeddings for offline beat tracking leads to a substantial improvement in F1 score (from 45.4\% to 73.3\%). However, due to the causality and real-time constraints, we do not employ these pre-trained speech models and Transformer architectures in this work. Instead, we used conventional spectral features and a CRNN architecture that can be computed in real-time.


\section{Conclusion}
\label{sec:conclusion}


This paper has introduced SingNet, a real-time beat and downbeat tracking system designed specifically for isolated singing voices. In addition to presenting SingNet, we have also proposed a dynamic particle filtering approach for inference, which leverages both instant predominant observations and offline inference results on historical data to improve the real-time beat and downbeat tracking accuracy. Specifically, we have proposed and compared three different methods for the dynamic particle filtering approach against the baseline BeatNet model. Evaluation results demonstrate significant improvement compared to the baseline. Overall, our proposed system provides a promising solution for real-time beat and downbeat tracking in singing voices, with potential applications in music information retrieval and interactive music systems.

\vfill\pagebreak

\bibliographystyle{IEEEbib-abbrev}

\bibliography{refs}

\begin{thebibliography}{10}

\bibitem{bock2016madmom}
S. B{\"o}ck et~al.,
\newblock ``Madmom: A new python audio and music signal processing library,''
\newblock in {\em Proc. of ACM Multimedia}, 2016, pp. 1174--1178.

\bibitem{krebs2016downbeat}
F. Krebs, S. B{\"o}ck, M. Dorfer, and G. Widmer,
\newblock ``Downbeat tracking using beat synchronous features with recurrent
  neural networks.,''
\newblock in {\em Proc. of ISMIR}, 2016, pp. 129--135.

\bibitem{fiocchi2018beat}
D. Fiocchi, M. Buccoli, M. Zanoni, F. Antonacci, and A. Sarti,
\newblock ``Beat tracking using recurrent neural network: a transfer learning
  approach,''
\newblock in {\em in 26th European Signal Processing Conference (EUSIPCO)}.
  IEEE, 2018, pp. 1915--1919.

\bibitem{chiu2021drum}
C.-Y. Chiu, A.~W.-Y. Su, and Y.-H. Yang,
\newblock ``Drum-aware ensemble architecture for improved joint musical beat
  and downbeat tracking,''
\newblock {\em IEEE Signal Processing Letters}, vol. 28, pp. 1100--1104, 2021.

\bibitem{di2021downbeat}
B. Di~Giorgi, M. Mauch, and M. Levy,
\newblock ``Downbeat tracking with tempo-invariant convolutional neural
  networks,''
\newblock {\em arXiv preprint arXiv:2102.02282}, 2021.

\bibitem{schreiber2018single}
H. Schreiber and M. M{\"u}ller,
\newblock ``A single-step approach to musical tempo estimation using a
  convolutional neural network.,''
\newblock in {\em Proc. of ISMIR}, 2018, pp. 98--105.

\bibitem{fuentes2019music}
M. Fuentes, B. McFee, H.~C. Crayencour, S. Essid, and J.~P. Bello,
\newblock ``A music structure informed downbeat tracking system using
  skip-chain conditional random fields and deep learning,''
\newblock in {\em in International Conference on Acoustics, Speech and Signal
  Processing (ICASSP)}. IEEE, 2019, pp. 481--485.

\bibitem{matthewdavies2019temporal}
M. Davies and S. B{\"o}ck,
\newblock ``Temporal convolutional networks for musical audio beat tracking,''
\newblock in {\em in 27th European Signal Processing Conference (EUSIPCO)}.
  IEEE, 2019, pp. 1--5.

\bibitem{hung2022modeling}
Y.-N. Hung, J.-C. Wang, X. Song, W.-T. Lu, and M. Won,
\newblock ``Modeling beats and downbeats with a time-frequency transformer,''
\newblock in {\em in International Conference on Acoustics, Speech and Signal
  Processing (ICASSP)}. IEEE, 2022, pp. 401--405.

\bibitem{zhao2022beat}
J. Zhao, G. Xia, and Y. Wang,
\newblock ``{Beat Transformer}: Demixed beat and downbeat tracking with dilated
  self-attention,''
\newblock {\em arXiv preprint arXiv:2209.07140}, 2022.

\bibitem{desblancs2022self}
D. Desblancs,
\newblock ``Self-supervised beat tracking in musical signals with polyphonic
  contrastive learning,''
\newblock {\em arXiv preprint arXiv:2201.01771}, 2022.

\bibitem{gkiokas2017convolutional}
A. Gkiokas and V. Katsouros,
\newblock ``Convolutional neural networks for real-time beat tracking: A
  dancing robot application.,''
\newblock in {\em Proc. of ISMIR}, 2017, pp. 286--293.

\bibitem{bock2011enhanced}
S. B{\"o}ck and M. Schedl,
\newblock ``Enhanced beat tracking with context-aware neural networks,''
\newblock in {\em Proc. of Int. Conf. Digital Audio Effects}, 2011, pp.
  135--139.

\bibitem{oliveira2010ibt}
J.~L. Oliveira, F. Gouyon, L.~G. Martins, and L.~P. Reis,
\newblock ``{IBT}: A real-time tempo and beat tracking system.,''
\newblock in {\em Proc. of ISMIR}, 2010, pp. 291--296.

\bibitem{heydari2021don}
M. Heydari and Z. Duan,
\newblock ``Don’t look back: An online beat tracking method using rnn and
  enhanced particle filtering,''
\newblock in {\em in International Conference on Acoustics, Speech and Signal
  Processing (ICASSP)}. IEEE, 2021, pp. 236--240.

\bibitem{heydari2021beatnet}
M. Heydari, F. Cwitkowitz, and Z. Duan,
\newblock ``{BeatNet}: {CRNN} and particle filtering for online joint beat,
  downbeat and meter tracking,''
\newblock in {\em Proc. of ISMIR}, 2021.

\bibitem{heydari2022singing}
M. Heydari and Z. Duan,
\newblock ``Singing beat tracking with self-supervised front-end and linear
  transformers,''
\newblock in {\em Proc. of ISMIR}, 2022.

\bibitem{chen2022wavlm}
S. Chen, C. Wang, Z. Chen, Y. Wu, S. Liu, Z. Chen, J. Li, N. Kanda, T.
  Yoshioka, X. Xiao, et~al.,
\newblock ``Wavlm: Large-scale self-supervised pre-training for full stack
  speech processing,''
\newblock {\em IEEE Journal of Selected Topics in Signal Processing}, 2022.

\bibitem{chang2022distilhubert}
H.-J. Chang, S.-w. Yang, and H.-y. Lee,
\newblock ``Distilhubert: Speech representation learning by layer-wise
  distillation of hidden-unit bert,''
\newblock in {\em in International Conference on Acoustics, Speech and Signal
  Processing (ICASSP)}. IEEE, 2022, pp. 7087--7091.

\bibitem{katharopoulos_et_al_2020}
A. Katharopoulos, A. Vyas, N. Pappas, and F. Fleuret,
\newblock ``Transformers are rnns: Fast autoregressive transformers with linear
  attention,''
\newblock in {\em Proc. of the International Conference on Machine Learning
  (ICML)}, 2020.

\bibitem{hainsworth2004particle}
S.~W. Hainsworth and M.~D. Macleod,
\newblock ``Particle filtering applied to musical tempo tracking,''
\newblock {\em EURASIP Journal on Advances in Signal Processing}, vol. 2004,
  no. 15, pp. 1--11, 2004.

\bibitem{krebs2015efficient}
F. Krebs, S. B{\"o}ck, and G. Widmer,
\newblock ``An efficient state-space model for joint tempo and meter
  tracking.,''
\newblock in {\em Proc. of ISMIR}, 2015, pp. 72--78.

\bibitem{rafii2018overview}
Z. Rafii, A. Liutkus, F.-R. St{\"o}ter, S.~I. Mimilakis, D. FitzGerald, and B.
  Pardo,
\newblock ``An overview of lead and accompaniment separation in music,''
\newblock {\em IEEE/ACM Transactions on Audio, Speech, and Language
  Processing}, vol. 26, no. 8, pp. 1307--1335, 2018.

\bibitem{marchand2015swing}
G. Tzanetakis, G. Essl, and P. Cook,
\newblock ``Automatic musical genre classification of audio signals,''
\newblock in {\em Proc. of ISMIR}, 2001.

\bibitem{defossez2021hybrid}
A. D{\'e}fossez,
\newblock ``Hybrid spectrogram and waveform source separation,''
\newblock {\em Proc. of ISMIR, Workshop on Music Source Separation}, 2021.

\bibitem{spijkervet_torchaudio_augmentations}
J. Spijkervet,
\newblock ``torchaudio-augmentations,''
\newblock 2021, Zenodo.

\end{thebibliography}

\end{document}